\documentclass{emulateapj} 
\usepackage{times}
\usepackage{booktabs}
\usepackage{times} 
\usepackage{amssymb}
\usepackage{amsmath}
\usepackage{natbib}
\usepackage{xspace}
\usepackage{multirow}
\usepackage{mathptmx} 
\usepackage{microtype}

\def \xmm{{\it XMM-Newton}\xspace}

\def \suzaku{{\it Suzaku}\xspace}
\def \chandra{{\it Chandra}\xspace}

\def \epicmos1{{\it EPIC}{\rm-MOS1}\xspace}
\def \epicmos2{{\it EPIC}{\rm-MOS2}\xspace}
\def \epicmos{{\it EPIC}{\rm-MOS}\xspace}

\def \nustar{{\it NuSTAR}\xspace}

\def \xselect{\textsc{xselect}\xspace}

\def \xisrmfgen{\textsc{xisrmfgen}\xspace}
\def \xissimarfgen{\textsc{xissimarfgen}\xspace}

\def \xstar{\textsc{xstar}\xspace}
\def \pl{\texttt{powerlaw}\xspace}

\def \tbabs{\texttt{tbabs}\xspace}
\def \bbody{\texttt{bbody}\xspace}

\def \pcfabs{\texttt{pcfabs}\xspace}

\def \zga{\texttt{zgauss}\xspace}

\def \logxi{\log(\xi/{\rm erg\,cm\,s}^{-1})}
\def \lognh{\log(N_{\rm H}/{\rm cm}^{-2})}
\def \vout{v_{\rm out}\xspace}
\def \nh{N_{\rm H}}
\def \nhgal{N_{\rm H,Gal}}

\def \lbol{L_{\rm bol}}
\def \mbh{M_{\rm BH}}
\def \rg{{$r_{\rm g}$}\xspace}

\def \dchidof{{\Delta\chi^{2}/\nu}\xspace}
\def \chidof{{\chi^{2}/\nu}\xspace}
\def \msun{{$M_{\odot}$}\xspace}
\def \pcmsq{{\rm cm$^{-2}$}\xspace}
\def \pcmcu{{\rm cm$^{-3}$}\xspace}
\def \flux{ {\rm erg\,cm$^{-2}$\,s$^{-1}$}\xspace }
\def \ergps{{\rm erg\,s$^{-1}$}\xspace}
\def \pout{\dot p_{\rm out}}
\def \pedd{\dot p_{\rm edd}}
\def \ergcmps{{\rm erg\,cm\,s$^{-1}$}\xspace}

\def \mout{\dot M_{\rm out}}

\def \msunyr{ {$M_{\odot}$\,yr$^{-1}$}\xspace }

\def \kms{ km\,s$^{-1}$ }
\def \fcov{ f_{\rm cov} }

\def \fexxv{Fe\,{\sc xxv}\xspace}
\def \fexxvi{Fe\,{\sc xxvi}\xspace}
\def \fexxvxxvi{Fe\,{\sc xxv-xxvi}\xspace}
\def \lya{Ly$\alpha$\xspace}
\def \hea{He$\alpha$\xspace}

\slugcomment{Accepted for publication in the ApJ}

\begin{document}
\title{Revealing the location and structure of the accretion disk-wind in PDS\,456}
\shorttitle{Disk-wind structure in PDS\,456}

\shortauthors{J.~Gofford et al.}
\author{J.~Gofford\altaffilmark{1},
		J.~N.~Reeves\altaffilmark{1,2},
		V.~Braito\altaffilmark{3},  
		E.~Nardini\altaffilmark{1}, 
		M.~T.~Costa\altaffilmark{1},  
		G.~A.~Matzeu\altaffilmark{1},  
		P.~O'Brien\altaffilmark{4}, 
		M.~Ward\altaffilmark{5},
		T.~J.~Turner\altaffilmark{2},
		L.~Miller\altaffilmark{6}
		}

\affil{
$^1$Astrophysics Group, School of Physical and Geographical Sciences, Keele 
University, Keele, Staffordshire, ST5 5BG, UK; j.a.gofford@keele.ac.uk \\
$^2$Center for Space Science and Technology, University of Maryland Baltimore County, 1000 Hilltop Circle, Baltimore, MD 21250, USA \\
$^3$INAF - Osservatorio Astronomico di Brera, Via Bianchi 46 I-23807 Merate (LC), Italy \\
$^4$Department of Physics and Astronomy, University of Leicester, University Road, Leicester, LE1 7RH, UK \\
$^5$Department of Physics, University of Durham, South Road, Durham, DH1 3LE, UK\\
$^6$Department of Physics, University of Oxford, Denys Wilkinson Building, Keble Road, Oxford, OX1 3RH, UK \\
}

\begin{abstract}
We present evidence for the rapid variability of the high velocity iron K-shell absorption in the nearby ($z=0.184$) quasar PDS\,456. From a recent long \suzaku observation in 2013 ($\sim1$\,Ms effective duration) we find that the the equivalent width of iron K absorption increases by a factor of $\sim5$ during the observation, increasing from $<105$\,eV within the first 100\,ks of the observation, towards a maximum depth of $\sim500$\,eV near the end. The implied outflow velocity of $\sim0.25$\,c is consistent with that claimed from earlier (2007, 2011) \suzaku observations. The absorption varies on time-scales as short as $\sim1$\,week. We show that this variability can be equally well attributed to either (i) an increase in column density, plausibly associated with a clumpy time-variable outflow, or (ii) the decreasing ionization of a smooth homogeneous outflow which is in photo-ionization equilibrium with the local photon field. The variability allows a direct measure of absorber location, which is constrained to within $r=200-3500$\,\rg of the black hole. Even in the most conservative case the kinetic power of the outflow is $\gtrsim6\%$ of the Eddington luminosity, with a mass outflow rate in excess of $\sim40\%$ of the Eddington accretion rate. The wind momentum rate is directly equivalent to the Eddington momentum rate which suggests that the flow may have been accelerated by continuum-scattering during an episode of Eddington-limited accretion.
\end{abstract}

\keywords{black hole physics --- quasars: individual (PDS 456) --- X-rays: galaxies}

\section{Introduction}
Outflows (or winds) are likely to be a natural and unavoidable result of the accretion process (e.g., \citealt{king03,ohsuga09}). In Active Galactic Nuclei (AGN) the feedback associated with outflowing matter is believed to play an important role in shaping the co-evolution of the central massive black hole and the host galaxy (\citealt{king03,dimatteo05}), plausibly leading to the observed AGN--host-galaxy relationships, e.g., $M-\sigma$ (\citealt{king10,zubovas12,mcquillin13}). Recently, a number of massive ($\nh\sim10^{23}$\,\pcmsq), high-velocity ($\vout\gtrsim0.1$\,c) outflows, as revealed through the presence of blue-shifted Fe\,K absorption at $E>7$\,keV in the rest-frame, have been found in luminous AGN (e.g., \citealt{pounds03, chartas03, reeves09, tombesi10, gofford13}). The large wind velocity---which indicates an origin directly associated with the accretion disk, hence leading them to be dubbed `disk-winds'---implies that the wind may be energetically significant in terms of feedback (i.e., $L/\lbol\sim0.5-5\%$, \citealt{hopkins10,dimatteo05}). 

Nonetheless, one key determination currently lacking is a direct measurement of the wind location with respect to the central black hole, which is of fundamental importance when it comes to determining the wind energetics. Previous studies of the disk-wind phenomenon have employed simple geometric and kinematic relations to constrain the location of the absorbing gas (e.g., see \citealt{tombesi12,tombesi13} for an outline of these arguments), but these arguments can lead to considerable uncertainties on the gas location which makes it difficult to confidently determine the wind energetics. The best way to overcome these limitations it to directly determine the location of the absorbing gas by establishing how it varies over short time-scales. However, whilst such line-variability has sometimes been observed in soft X-ray grating spectra below $E=2-3$\,keV, hence leading to robust constraints on the location of the soft X-ray warm absorber in a number of AGN, e.g., in MRK\,509 (\citealt{kaastra12}) and MR\,2251-178 (\citealt{reeves13a}), to date it has proven much more difficult to measure the necessary line variability at harder X-ray energies (i.e., $E>7$\,keV) where modern detectors tend to be less sensitive (e.g., \citealt{giustini11}). In this work we overcome these limitations and report the first direct constraints on the location of a high-velocity Fe\,K disk-wind, as measured in the powerful quasar PDS\,456. 

PDS\,456 ($z=0.184$) is a luminous ($\lbol=10^{47}$\,\ergps; \citealt{simpson99,reeves00}) radio-quiet quasar which harbours one of the most powerful Fe\,K disk-winds currently known (\citealt{reeves09}, hereafter `R09'). \xmm first found the X-ray spectrum to be absorbed at $E>7$\,keV in 2001, with the absorption attributable to highly-ionised iron (\citealt{reeves03}). A subsequent \suzaku observation in 2007 revealed two highly significant absorption lines at $9.08$ and $9.66$\,keV in the quasar rest-frame (R09). These lines are most likely associated with resonant absorption from \fexxv~\hea and \fexxvi~\lya, which hence implies an outflow velocity in the range $\vout\sim0.25-0.3$\,c (R09). In a more recent \suzaku observation (2011) we again found a broad absorption trough at $\sim9$\,keV in the source rest-frame (Reeves et al. 2013, hereafter `R13'), and we found that the absorption in both 2007 and 2011 could be due to the same flow of gas which is in photo-ionization equilibrium with the emergent X-ray emission. However, due to the large time difference between the observations, direct constraints could not be placed on the radial location of the absorbing gas. Even so, we were able to describe the absorption in both observations using the self-consistent disk-wind models of \citet{sim08,sim10a}, hence showing the variable wind profile to be consistent with a radiatively-driven flow launched from the inner accretion disk (R13).

In this paper we report the first results from our extensive observational campaign of PDS\,456 with the \suzaku, \xmm and \nustar satellites in 2013 (Feb--Sept). Here, we focus specifically on characterising the remarkable spectral variability exhibited by the Fe\,K wind during the new long \suzaku observation (Feb--Mar 2013), and use this variability to place the first direct constraints on the wind location. A full spectral analysis of both these \suzaku data and the accompanying contemporaneous \xmm and \nustar campaign will be presented in forthcoming work.

\section{Data reduction}
\suzaku (\citealt{mitsuda07}) observed PDS\,456 at the aim-point of the X-ray Imaging Spectrometers (XIS; \citealt{koyama07}). Due to scheduling constraints the observation comprised three sequences: the first (OBSID: 707035010, hereafter 2013a) has a duration of $\sim$441\,ks and was obtained between 21--26 Feb 2013, whilst the second (OBSID: 707035020, hereafter 2013b) and third (OBSID: 707035030; hereafter 2013c) were obtained consecutively between 3--11 March 2013 and have durations of $\sim$404\,ks and $\sim$245\,ks, respectively. The effective duration of the campaign was $\sim1$\,Ms. A detailed summary of the three sequences is given in Table~\ref{tab:observation_details}. Analogous parameters for the 2007 and 2011 \suzaku observations are also noted for comparison. 

\begin{deluxetable*}{lccccc}
\tabletypesize{\footnotesize}
\tablecolumns{6}
\tablewidth{0pt}
\tablecaption{Summary of PDS\,456 observations with \suzaku\label{tab:observation_details}}
\tablehead{
\colhead{Parameter} &
\colhead{2007} &
\colhead{2011} &
\colhead{2013a} & 
\colhead{2013b} & 
\colhead{2013c}
}
\startdata
Sequence Number & 701056010 & 705041010 & 707035010 & 707035020 & 707035030 \\[0.5ex]
Start Date, Time (UT) & 2007-02-24, 17:58 & 2011-03-16, 15:00 & 2013-02-21, 21:22 & 2013-03-03, 19:43 & 2013-03-08, 12:00 \\[0.5ex]
End Date, Time (UT) & 2007-03-01, 00:51 & 2011-03-19, 08:33 & 2013-02-26, 23:51 & 2013-03-08, 12:00 & 2013-03-11, 09:00 \\[0.5ex]
Duration (ks)$^{a}$ & -- & -- & $0-440.914$ & $858.026-1262.252$ & $1262.253-1510.654$\\[0.5ex]
Exposure (ks) & 190.6 & 125.6 & 182.3 & 164.8 & 108.3\\[0.5ex]
Count Rate ($10^{-2}$\,ct\,s$^{-1}$) & $27.22\pm0.09$ & $14.1\pm0.08$ & $6.74\pm0.04$ & $4.35\pm0.04$ & $5.15\pm0.05$\\
\midrule
$F({\rm 0.5-2\,keV})^{b}$ & 3.46 & 1.36 & 0.59 & 0.29 & 0.45 \\[0.5ex]
$F({\rm 2-10\,keV})^{b}$ & 3.55 & 2.84 & 2.07 & 1.52 & 1.51
\enddata
\tablenotetext{a}{time-splits for the 2013 sequences}
\tablenotetext{b}{continuum flux, in units of $\times10^{-12}$\flux}
\end{deluxetable*}

Following the process outlined in the \suzaku data reduction guide\footnote{http://heasarc.gsfc.nasa.gov/docs/suzaku/analysis/abc/‎} we extract spectral data for the functioning XIS(0,\,1,\,3) CCDs in each sequence using {\sc heasoft} (v6.14) and the latest version of the calibration database (Aug 2013). Data were selected from the $3\times3$ and $5\times5$ edit-modes and then processed according to the recommended screening criteria. XIS source products were selected from circular regions $1.5^{\prime}$ in radius, with the background contribution estimated from four offset circular regions of equal radius. Spectra and light-curves were extracted from the cleaned event files using \xselect, with the response matrices ({\sc rmf}) and ancillary response files ({\sc arf}) for each detector created using the \xisrmfgen and \xissimarfgen tasks. After verifying their consistency we created an XIS-FI spectrum for each sequence by combining the spectra and response files for each set of front-illuminated (FI) XIS\,0 and 3 CCDs; the net on-source exposures were 182.3\,ks, 164.8\,ks and 108.3\,ks for sequences 2013a, 2013b and 2013c, respectively, culminating in a net \suzaku exposure of 455.4\,ks for PDS\,456 during the 2013 observation.

Unfortunately, PDS\,456 is not detected by the Hard X-ray Detector (\citealt{takahashi07}). However, a definitive hard X-ray spectrum has since been obtained by \nustar in Sept~2013. Details of these observations are deferred to future work.
 
\section{Spectral Analysis}
In this work we focus our attention on the data obtained by the FI XIS CCDs because they provide the highest effective area and lowest background rate in the crucial Fe\,K band. For analysis, we grouped each of the XIS-FI spectra to the approximate half-width at half-maximum (HWHM) energy resolution of the XIS (i.e., $\sim$60\,eV at 6\,keV), and adopted an additional minimum grouping of 40 counts per energy bin such that the $\chi^{2}$ minimization technique could be used during spectral fitting. We consider the XIS-FI data between $0.6-10$\,keV when fitting the data, ignoring between $1.7-1.9$\,keV due to uncertainties with the XIS Si detector edge. Unless otherwise stated, all statistics are given relative to the Fe\,K band (i.e., between $5-10$\,keV) to ensure that they are driven by the Fe\,K absorption rather than the soft X-ray data which typically dominates the statistics in XIS data. Parameter errors are quoted at the 90\% confidence interval (corresponding to $\Delta\chi^{2}=2.71$ for 1 parameter of interest), whilst the standard 1$\sigma$ error bars are displayed in all plotted spectra. We adopt values of $H_{\rm 0}=70$\,km\,s$^{-1}$\,Mpc$^{-1}$, and $\Omega_{\Lambda_{\rm 0}}=0.73$ throughout.
 
\begin{figure}
\epsscale{1}
\plotone{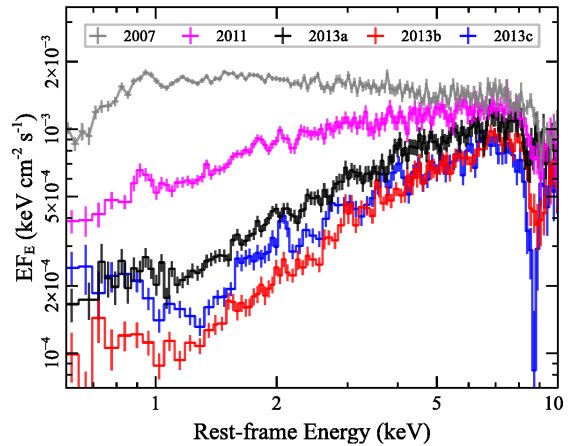}
\caption{
Comparison of spectra from all \suzaku observations of PDS\,456 to date (2007, 2011, 2013a--c). The 2007 observation (grey spectrum) is the brightest, with a roughly power-law spectral shape, whilst the 2011 and 2013 observations get progressively harder and more absorbed. Absorption is persistent at $\sim9$\,keV in all of the \suzaku observations, and is especially prominent during the 2013 observation.}
\label{fig:eeuf_comparison}
\end{figure}

\subsection{Individual sequences}
\begin{figure*}
\epsscale{0.5}
\plotone{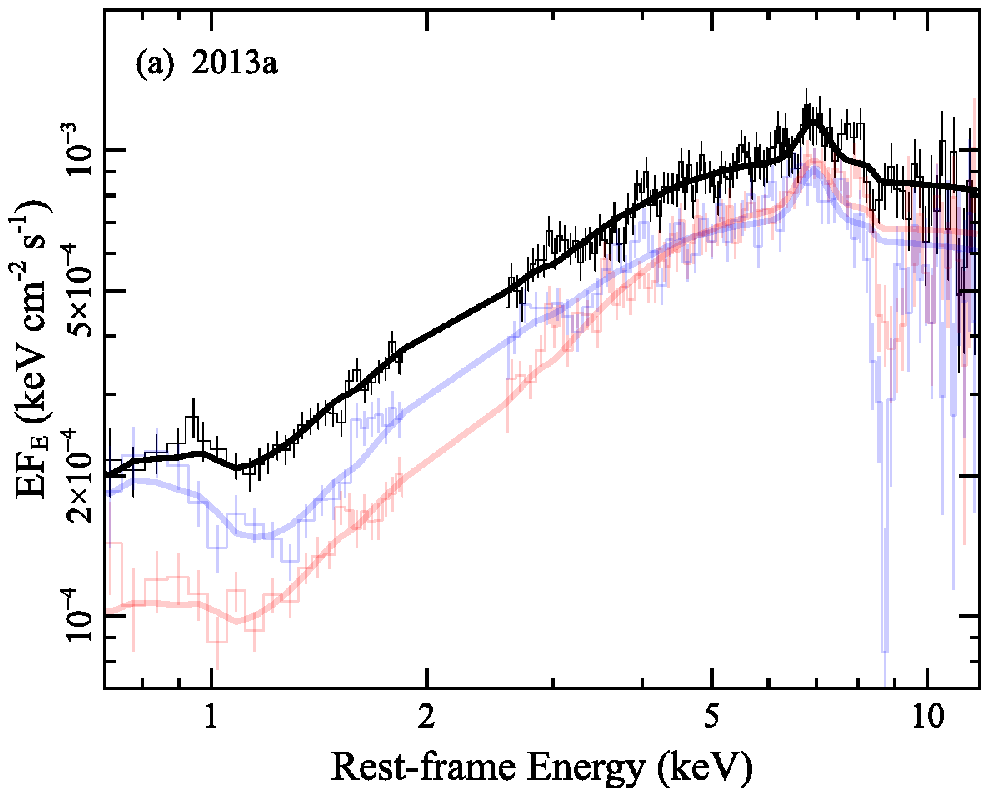}\hspace{0.4cm}
\plotone{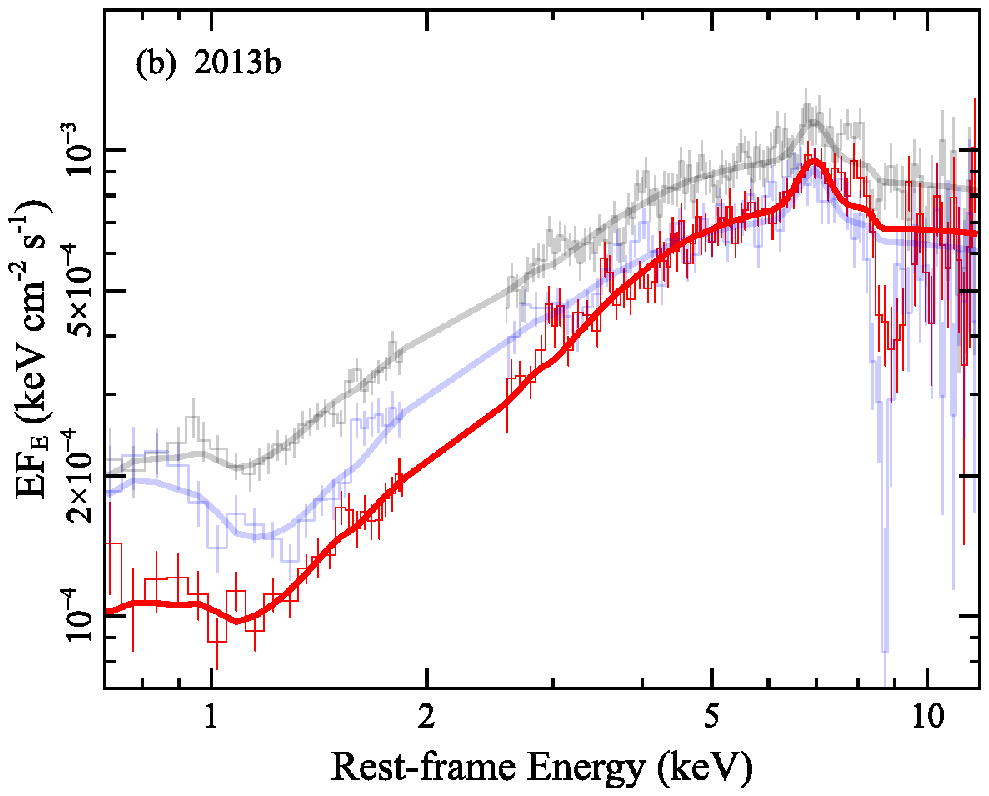}
\vspace{0.4cm}
\plotone{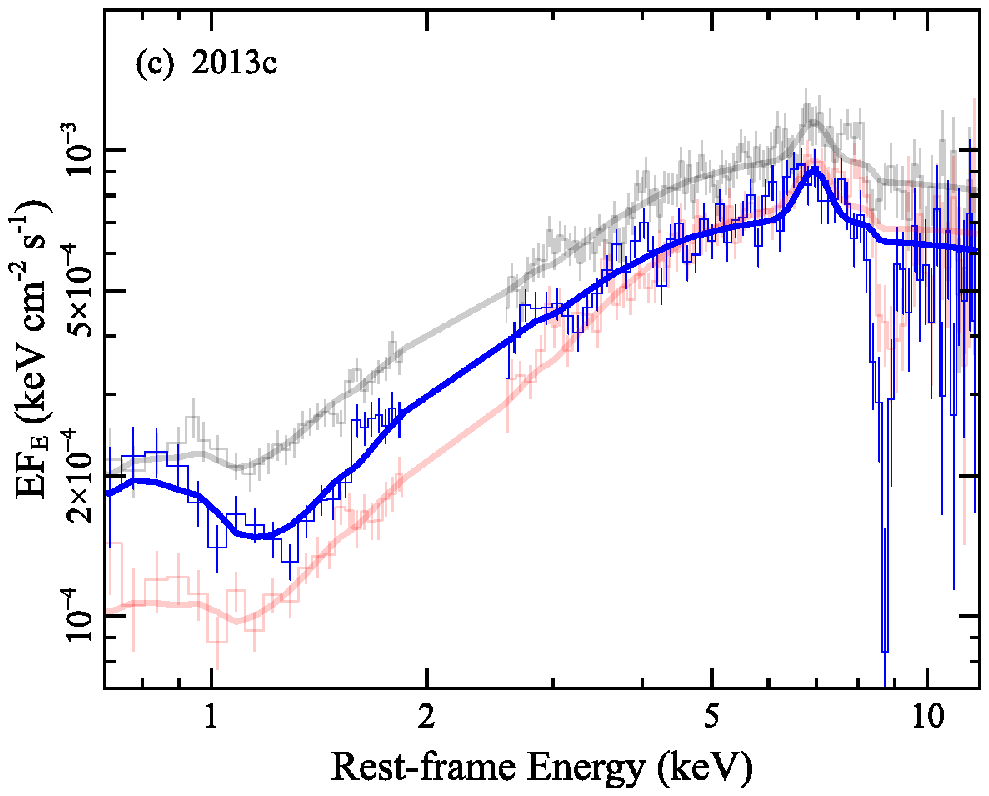}\hspace{0.4cm}
\plotone{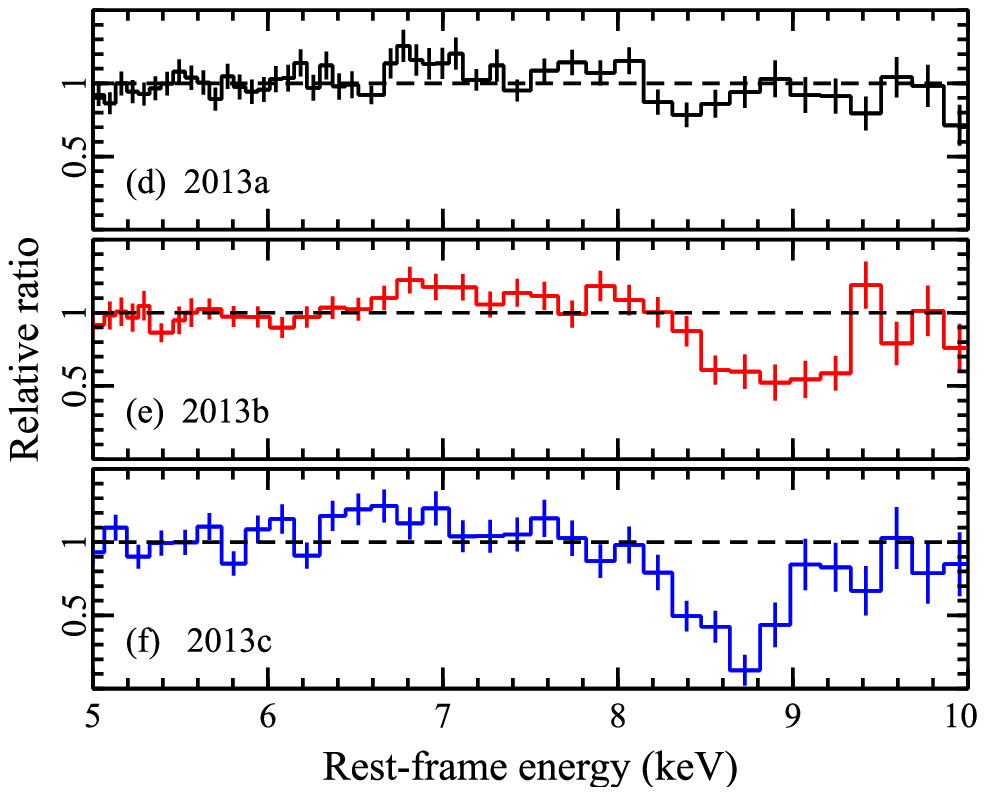}
\caption{
Fluxed rest-frame spectra from the three 2013 \suzaku sequences. Panel (a)--(c) shows the unfolded spectra for each of the 2013 sequences and their corresponding best-fit continuum model. Sequences 2013a, 2013b and 2013c are shown by the black, red, and blue spectra, respectively. Note that these fluxed spectra have been created by unfolding the data and model against a $\Gamma=2$ power-law, with the best-fitting continuum model overlaid afterwards; they have not been unfolded against the best-fit model including an absorption line. Panels (d)--(f) show the ratio spectra of the three 2013 sequences against the baseline partially-covered continuum model described in the text. Note the strong increase in depth of the Fe\,K absorption from sequence 2013a to 2013c. Data have been re-binned to the approximate FWHM resolution of the XIS-FI for clarity.}
\label{fig:eeuf_sequences}
\end{figure*}

We began our analysis by considering the XIS-FI spectra from each sequence individually. Figure~\ref{fig:eeuf_comparison} shows a raw comparison between all of the \suzaku observations of PDS\,456 taken to date (2007, 2011, 2013a--c). The data are unfolded against a simple $\Gamma=2$ power-law model and uncorrected for Galactic absorption. It is apparent from Figure~\ref{fig:eeuf_comparison} that all of the 2013 sequences are reasonably hard ($\Gamma=1.27$ between $3-5$\,keV, compared to $\Gamma=2.3-2.4$ in 2007), with a spectral shape reminiscent of (but still harder than) that seen in the absorbed 2011 observation (R13). The soft X-ray flux is much lower in 2013 than in the relatively unabsorbed 2007 observation (R09, R13), with a mean $\langle F_{\rm 0.5-2} \rangle = 4.43\times10^{-13}$\flux in 2013 compared to $3.46\times10^{-12}$\flux in 2007, whilst $\langle F_{\rm 2-10} \rangle $ is also lower (see Table~\ref{tab:observation_details}). These factors suggest that PDS\,456 is again heavily absorbed during the 2013 \suzaku epoch. This is further alluded to by the deep Fe\,K absorption trough at $\sim9$\,keV (rest-frame) in sequences 2013b and 2013c (see Figure~\ref{fig:eeuf_sequences}). The remainder of this work focusses on characterising this absorption in greater detail.

%FIGURE ONE - two column mode

In order to probe the absorption system it was first necessary to parametrise the continuum. We find that a partially-covered absorption model provides a good simultaneous fit to the three 2013 sequences. In summary, and noting that a fully detailed spectral de-composition of these data is to be presented in subsequent work, we parametrise the continuum with a phenomenological model of the form $\tbabs(\pcfabs_{1}\times\pcfabs_{2}\times\pl+\bbody+\zga)$, where $\tbabs$ (\citealt{wilms00}) accounts for the Galactic absorption column density of $N_{\rm H,Gal}=2\times10^{21}$\,\pcmsq, $\pl$ is the underlying power-law continuum, $\pcfabs_{1,2}$ are two neutral partially-covering absorbers, and the $\bbody$ is an ad-hoc black-body with $kT\sim(106\pm5)$\,eV that provides some necessary soft emission at $E\lesssim1.5$\,keV. This soft component may be associated with the reprocessed emission from the outer wind, e.g., similar to what is seen in some wind-dominated Ultra-Luminous X-ray (ULX) sources (\citealt{middleton14}), with the intrinsic emission of the accretion disc, or have another origin. For simplicity, we opt to not absorb the \bbody component with the partially-covering gas to prevent degeneracies between its normalization and the gas covering fraction. We note, however, that an equivalent description of the data can be achieved if the \bbody is absorbed; the ensued parameters for the Fe\,K absorption are also unaffected by how the soft component is modelled. A full discussion regarding the origin of the soft component is deferred to future work.

The $\zga$ component in our baseline model corresponds to the broad ionised emission line at $\sim6.9$\,keV (rest-frame), which might be associated with light scattered from the disk-wind (see R13 for a discussion); we fit this broad profile with a (fixed) width of $\sigma=400$\,eV, as found for the 2011 observation by R13. We adopt assume a (fixed) power-law slope of $\Gamma=2.4$ to ensure consistency with both the bright (unabsorbed) 2007 \suzaku observation (R09) and the new \nustar hard X-ray spectrum (which will be presented in subsequent work). Allowing $\Gamma$ free to vary from this value does not result in a statistical improvement to the fit. We account for the changes in intra-sequence curvature by allowing the covering fraction ($f_{\rm cov}$) of the partially-covering absorbers to vary freely; covering fractions of $f_{\rm cov,1}\simeq0.75-0.95$ and $f_{\rm cov,2}\simeq0.45-0.75$, for tied column densities of $\log N_{\rm H,1}=21.9\pm0.1$ and $\log N_{\rm H,2}=22.9\pm0.1$, respectively, are sufficient to model all three sequences. Note that whilst we do not discuss the physical nature of the partially-covering gas in detail here, it could plausibly be associated with either clumps within a disk-wind (e.g., R13) or possibly with clouds in the more extended Broad Line Region (BLR) (e.g., as has been suggested in NGC\,1365, \citealt{risaliti09}; Mrk\,766, \citealt{risaliti11} and MR\,2251-178, \citealt{reeves13a}). Full parameters for the baseline model are noted in Table~\ref{tab:baseline_model}.

In general, this simple phenomenological model is consistent with that employed by R13 to jointly describe the 2007 and 2011 observations, and we find that it again provides a good account of the spectral shape in 2013. The full fit statistic between $0.6-10$\,keV is $\chidof=716.2/505$, with the strong absorption in the Fe\,K band being the main deviation from the continuum model (see Figure~\ref{fig:eeuf_sequences}, panels d--f). Figure~\ref{fig:eeuf_sequences}(a--c) shows the fluxed 2013 sequences and their best-fit model. Note that obtaining an accurate description of the spectral variability during 2013 appears to be contingent on the source being obscured by two layers of partially-covering gas. This is consistent with what was found by R13 from the analysis of the previous 2007 and 2011 \suzaku spectra.

\begin{deluxetable}{llccc}
\tabletypesize{\footnotesize}
\tablecolumns{5}
\tablewidth{0pt}
\tablecaption{Baseline continuum parameters\label{tab:baseline_model}}
\tablehead{
\colhead{Component} &
\colhead{Parameter} &
\colhead{2013a} & 
\colhead{2013b} & 
\colhead{2013c}
}
\startdata
$\tbabs$ 		& $\nhgal$ 	 	& \multicolumn{3}{c}{$2\times10^{21}$\,cm$^{-2}$}\\[0.5ex]
$\pl$ 			& $\Gamma$ 	 	& $2.4^{\ast}$ & -- & -- \\[0.5ex]
		  		& norm$^{a}$ 	& $2.14\pm0.06$ & $1.69\pm0.05$ & $1.45\pm0.05$\\[0.5ex]
$\pcfabs_{1}$ 	& $\log\nh^{b}$ & $21.9\pm0.1$ & $21.9^{t}$ & $21.9^{t}$\\[0.5ex]
		  		& $\fcov$ (\%) 	& $77\pm3$ & $86\pm4$ & $>95$\\[0.5ex]
$\pcfabs_{2}$ 	& $\log\nh^{b}$ & $22.9\pm0.1$ & $22.9^{t}$ & $22.9^{t}$\\[0.5ex]
		  		& $\fcov$ (\%) 	& $63\pm3$ & $73\pm2$ & $50\pm4$\\[0.5ex]
$\bbody$ 		& $kT$ (eV) 	& $106\pm5$ & $106^{t}$ & $106^{t}$\\[0.5ex]
				& norm$^{c}$ 	& $1.7\pm0.3$ & $0.9\pm0.2$ & $2.4\pm0.5$\\[0.5ex]
$\zga$ 			& $E$ (keV)		& $6.94\pm0.10$ & $6.94^{t}$ & $6.94^{t}$\\[0.5ex]
				& $\sigma$ (eV) & $400^{\ast}$ & -- & --\\[0.5ex]
				& EW (eV)		& $220\pm30$ & $290\pm40$ & $330\pm40$

\enddata
\tablenotetext{a}{power-law normalisation, in units of $10^{-3}$\,ph\,keV$^{-1}$\,cm$^{-2}$\,s$^{-1}$}
\tablenotetext{b}{Logarithm of the absorber column density. Unit of $\nh$ is cm$^{-2}$}
\tablenotetext{c}{black-body normalisation in units of $10^{-5}(L_{39}/D^{2}_{10})$, where $L_{39}$ is the source luminosity in units of $10^{39}$\,erg\,s$^{-1}$ and $D_{10}$ is the distance to the source in units of $10$\,kpc.}
\tablenotetext{$\ast$}{indicates a parameter fixed during fitting,}
\tablenotetext{t}{parameter tied during fitting.}
\end{deluxetable}

It is clear that despite the baseline model providing a reasonable fit to the continuum the fit statistic between $5-10$\,keV (rest-frame) is still poor due to the strong absorption ($\chidof=383.2/190$ over this energy range, $P_{\rm null}=4.1\times10^{-15}$). The ratio spectra (see Figure~\ref{fig:eeuf_sequences}, d--f) demonstrate that the depth of absorption increases across the three sequences, being weakest in the 2013a sequence and strongest in 2013c. Modeling the profile with a Gaussian (see Table~\ref{tab:sequences} for parameters) indicates that, for a best-fit variable width of $\sigma=(250\pm60)$\,eV which was left tied between the sequences, the equivalent width ($EW$) of absorption increases by a factor of $\sim5$ throughout the observation, from $<110$ in 2013a to $(486^{+71}_{-69})$\,eV in 2013c. The line centroid also appears to be at a lower energy in 2013c than in 2013b at the 90\% level, decreasing from $(8.80\pm0.10)$ to $(8.65\pm0.06)$\,keV, indicative of a possible change in outflow velocity or ionization state. The Gaussian parametrisation results in a $\dchidof=-159.2/6$ improvement with respect to the baseline continuum, for an overall fit statistic of $\chidof=224.0/184$ in the Fe\,K band.

Having parametrised the profile, we replaced the Gaussian with a custom generated \xstar (v2.21bn13, \citealt{bautista01}) absorption table with a $\Gamma=2.4$ illuminating continuum and a turbulent line broadening of $v_{\rm turb}=5000$\,km\,s$^{-1}$. This value for $v_{\rm turb}$ was adopted as it provided the best description of the broadness of the absorption profile. We also tried other grids but found that in grids with $v_{\rm turb}=(1000,\,3000)$\,km\,s$^{-1}$ the saturated too quickly to fit the observed breadth and equivalent width of the profile, whilst $v_{\rm turb}=10000$\,km\,s$^{-1}$ was simply too broad to provide a satisfactory fit to the data. Using the $v_{\rm turb}=5000$\,km\,s$^{-1}$ absorption table we thus investigated two scenarios that could give rise to the observed line variability. In the first (Model A), we tied the column density ($\nh$) between the sequences but allowed the ionization parameter ($\xi$) to vary. This scenario corresponds to an absorber that remains persistently in the line-of-sight (LOS) but whose ionization changes (decreases) during the observation. In the second scenario (Model B), we allowed the $\nh$ to vary but tied the $\xi$. This scenario is intended to model the case where an inhomogeneous absorber crosses the LOS during the observation, with $\nh$ subsequently increasing. The outflow velocity $\vout$ was initially allowed to vary in both cases. 

\begin{deluxetable}{llccc}
\tabletypesize{\footnotesize}
\tablecolumns{5}
\tablewidth{0pt}
\tablecaption{Iron K absorption parameters for 2013 sequences.\label{tab:sequences}}
\tablehead{
&
\colhead{Parameter} & 
\colhead{2013a} & 
\colhead{2013b} & 
\colhead{2013c}
}
\startdata
& \textbf{\underline{Gaussian}} & & & \\[0.5ex]
(1) & Line Energy\,(keV) & $8.88^{t}$ & $8.88\pm0.10$ & $8.65\pm0.06$\\[0.5ex]
(2) & $\sigma$-width (eV) & $250^{t}$ & $250\pm60$ & $250^{t}$ \\[0.5ex]
(3) & EW\,(eV) & $<110$ & $310\pm80$ & $490\pm70$\\[0.5ex]
(4) & $\dchidof$ & -- & $-42.1/3$ & $-129.6/3$ \\[0.5ex]
\midrule
 & \textbf{\underline{\xstar}} & & & \\[0.5ex]
(5) & $\logxi$ & $>4.35$ & $3.8\pm0.1$ & $3.4\pm0.1$\\[0.5ex]
(6) & $\lognh$ & $<22.9$ & $23.4\pm0.2$ & $23.9\pm0.1$\\[0.5ex]
(7) & $\vout/c$ & -- & $-0.244\pm0.008$ & $-0.227\pm0.007$

\enddata
\tablecomments{(1) Rest-frame energy of Gaussian Absorption line; (2) $\sigma$-width of Gaussian; (3) Equivalent width of Gaussian; (4) change in $\chidof$ upon adding a Gaussian line to the baseline continuum model. A negative value indicates a statistical improvement; (5) \xstar parameters for Model A, in which $\logxi$ is allowed to vary for a constant $\lognh=23.6^{+0.1}_{-0.2}$; (6) \xstar parameters for Model B, where $\nh$ is allowed to vary at a constant $\logxi=3.48\pm0.14$; (7) inferred outflow velocity in the quasar rest-frame ($z=0.184$). Negative values denote a net blue-shift}
\tablenotetext{t}{indicates that a parameter was tied during fitting.}
\end{deluxetable}

\begin{figure}
	\plotone{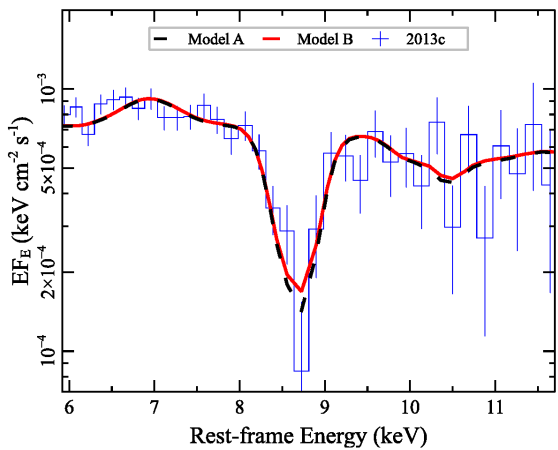}
	\caption{Fluxed rest-frame spectrum from sequence 2013c overlaid with the best-fit \xstar models for the deep absorption line. The fit for models A and B, which, as described in the text, provide statistically equivalent descriptions of the observed profile, are shown by the black (dashed) and red (solid) lines, respectively. As before, the fluxed spectra has been unfolded against a reference $\Gamma=2$ power-law continuum with the model overlaid afterwards; they have not been unfolded against the best-fitting model.}
	\label{fig:model_ratio}
\end{figure}

The parameters for both of these models are listed in Table~\ref{tab:sequences}. In both cases, the outflow velocity of the absorber appearing to be slightly lower in 2013c than in 2013b at the $P_{\text{f-test}}>99\%$ level, which mirrors the subtle change in centroid energy found during Gaussian fitting (see Table~\ref{tab:sequences}). Statistically, the two models yield equivalently good fits to the Fe\,K band, with $\chidof=211.2/184$ and $\chidof=208.9/184$ for Model A and Model B, respectively. A comparison of the two fits, as applied to sequence 2013c, is shown in Figure~\ref{fig:model_ratio}. It is evident that both models give a good description of the data from the time-averaged sequences.

\vspace{0.3cm}

\subsection{Time-sliced spectra}
\label{sub:time-sliced_spectra}
\begin{figure}[t]
\epsscale{1}
\plotone{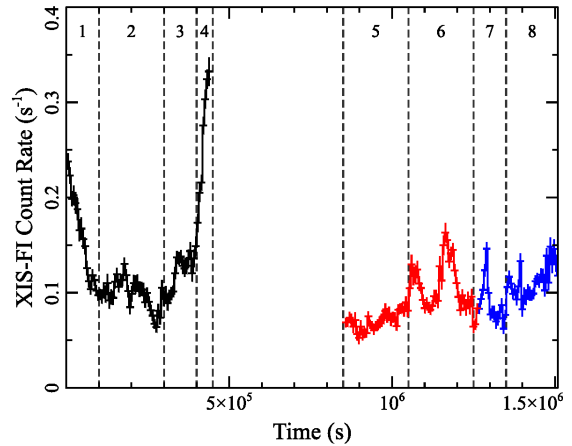}
\caption{
Combined XIS\,03 0.5--10\,keV light-curves for the 2013 \suzaku data, binned to orbital (5760\,s) bins. Light-curves for sequences 2013a, 2013b and 2013c are shown by the black, red, and blue data points, respectively. The gap in the coverage corresponds to the one week scheduling gap between 2013a and 2013b. The vertical dotted lines mark boundaries of the spectral slices $1-8$ as described in the text. Note the strong flare between $400-450$\,ks (slice $4$) at the end of 2013a.}
\label{fig:lc_slices}
\end{figure}

The composite light-curve (see Figure~\ref{fig:lc_slices}) indicates that the X-ray flux for PDS\,456 is variable during the 2013 \suzaku observation. This is most apparent in Sequence 2013a, which has a strong $\times3-4$ flare in flux between $400-450$\,ks. Smaller flares ($\times1.5-2$) are also evident in the latter half of sequence 2013b, but in general the later two sequences appear to be in a state of relatively constant flux. Unfortunately, the scheduling gap between sequences 2013a and 2013b occurs during the large flare, meaning that we are unable to fully trace how it evolves with time. Even so, the remaining data are all of sufficient quality to enable a time-resolved analysis of the absorption profile, thereby sampling its variability over shorter time-scales. Guided by the visual properties of the light-curve we thus split the spectrum into a total of 8 slices; these are overlaid on the light-curve in Figure~\ref{fig:lc_slices}. Slices 1 and 2 trace the decline and subsequent quiescent period in the first half of sequence 2013a, whilst slices 3 and 4 trace both the initial onset of the flare and the flare itself, respectively. Slices 5--8 then split sequences 2013b and 2013c roughly in half, avoiding any smaller flares. The timing periods for each slice are noted in Table~\ref{tab:slices}.

\begin{figure}[t] 
\epsscale{1}
\plotone{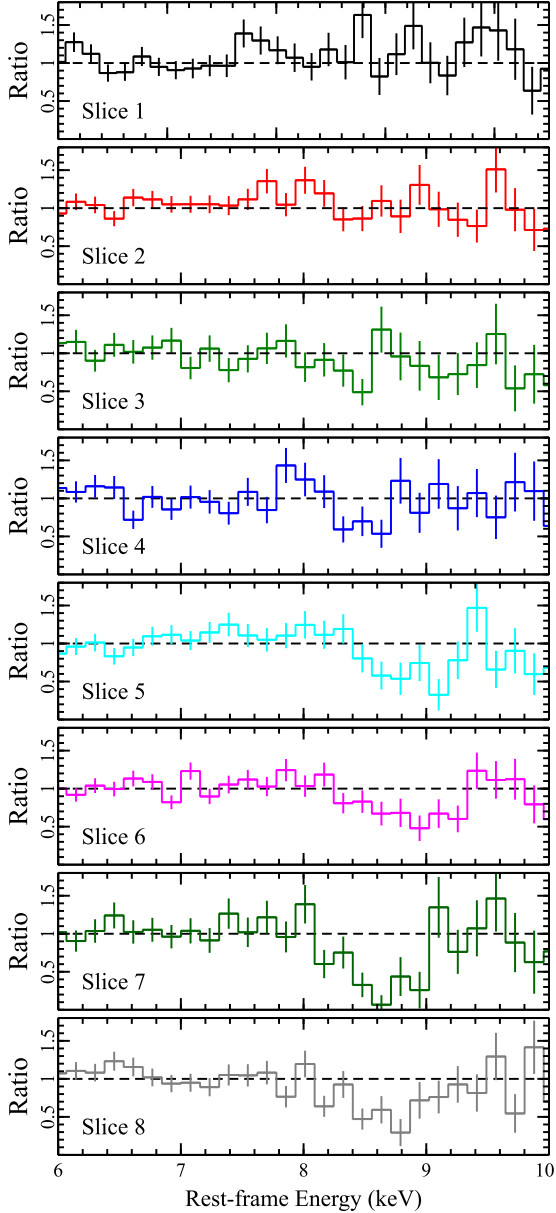}
\caption{
Ratio spectra of the \suzaku slices in the Fe\,K region, as defined in Figure\,2. The iron K absorption feature becomes progressively stronger through the observation, reaching a maximum depth during slice $G$ near the end of the observation. Best fit parameters for the absorption are listed in Table\,2.}
\label{fig:slices_ratio} 
\end{figure}

We thus re-sampled the properties of the absorption using the time-sliced spectra, first with a simple Gaussian and then with \xstar. The baseline model (as described earlier) again provides a good description of all 8 slices; the rest-frame $6-10$\,keV ratio spectra with respect to this model are shown in Figure~\ref{fig:slices_ratio}. In general, the time-sliced results are consistent with those for the time-averaged sequences: Fe\,K absorption is not statistically required at the start of the observation (slices $1-3$), with the first tentative detection (i.e., at $\geq90\%$ confidence) occurring in slice $4$. The absorption then gets sequentially stronger through slices $5$ and $6$, before reaching an eventual maximum depth (and highest significance of $\dchidof=-54.5/3$) in slice $7$. In slice $8$ the line seems to get slightly less significant, but its overall parameters are consistent with those found in slice $7$. For a common line width of $\sigma=(240\pm70)$\,eV, the $EW$ of the profile increases from $<105$ to $(480\pm125)$\,eV in the $\sim1$\,Ms between slices $1$ and $7$, i.e., by a factor $\sim5$. The line centroid is again found to be $\sim8.6-8.9$\,keV in the quasar rest-frame, consistent with what was found before in the time-averaged sequences.

We then tested \xstar models A and B from before, initially with the absorbers $\vout$ left free to vary (except in slices $1-4$, where it was tied between the slices due to the non-detections at the start of the observation). We again find that both models provide acceptable descriptions of the data, with $\chidof=358.2/351$ and $346.2/351$ for models A and B, respectively. The outflow velocity is in the range $\vout\sim0.24-0.27$\,c, which is consistent with previous analyses (R09, R13). Tying $\vout$ between the slices worsens the fit in both cases, yielding $\chidof=362.2/356$ and $\chidof=362.1/356$ for model A and B. Full model parameters are given in Table~\ref{tab:slices}. Overall, these results suggest that the line variability can be well described through a transiting inhomogeneous cloud or by ionization changes in a homogeneous absorber. 

\begin{deluxetable*}{clccccccc}
\tabletypesize{\footnotesize}
\tablecaption{Iron K absorption parameters from the spectral slices.\label{tab:slices}}
\tablewidth{0pt}
\tablehead{
\colhead{\multirow{2}{*}{Slice}} & 
\colhead{\multirow{2}{*}{Time (ks)}} & 
\colhead{Energy (keV)} & 
\colhead{EW (eV)} & 
\colhead{$N_{\rm H}$} & 
\colhead{$\log\xi$} & 
\colhead{$v_{\rm out}/c$} & 
\colhead{Flux} &
\colhead{$\dchidof$}\\[0.5ex]
 &
 &
\colhead{(1)} &
\colhead{(2)} &
\colhead{(3)} &
\colhead{(4)} &
\colhead{(5)} &
\colhead{(6)} &
\colhead{(7)}
}
\startdata

% & TIME & ENERGY & EW & NH & logxi & vout & FLUX
1 & $0-100$ & $8.50^{t}$ & $<105$ & $<22.6$ & $>4.28$ & $-0.22^{t}$ & 3.00 (6.70) & NR \\[0.5ex]
2 & $100-300$ & $8.50^{t}$ & $<132$ & $<22.8$ & $>3.97$ & $-0.22^{t}$ & 1.94 (4.86) & NR\\[0.5ex]
3 & $300-400$ & $8.50^{t}$ & $<257$ & $<23.5$ & $>3.28$ & $-0.22^{t}$ & 2.20 (5.05) & NR \\[0.5ex]
4 & $400-450$ & $8.50\pm0.15$ & $220\pm100$ & $23.0\pm0.4$ & $3.95\pm0.30$ & $-0.22\pm0.02$ & 4.72 (11.81) & $-8.2/3$ \\[0.5ex]
5 & $850-1050$ & $8.90\pm0.16$ & $300\pm120$ & $23.4\pm0.2$ & $3.66\pm0.20$ & $-0.26\pm0.02$& 1.46 (4.15) & $-13.2/3$ \\[0.5ex]
6 & $1050-1250$ & $8.92\pm0.14$ & $300\pm95$ & $23.5\pm0.2$ & $3.62\pm0.14$ & $-0.27\pm0.01$ & 2.14 (5.19) & $-24.9/3$\\[0.5ex]
7 & $1250-1350$ & $8.63\pm0.13$ & $480\pm125$ & $23.8\pm0.2$ & $3.49^{+0.07}_{-0.03}$ & $-0.24\pm0.01$ & 1.70 (4.46) & $-54.5/3$\\[0.5ex]
8 & $1350-1510$ & $8.70\pm0.11$ & $360\pm100$ & $23.7\pm0.1$ & $3.31\pm0.03$ & $-0.25\pm0.01$ & 2.05 (4.88) & $-30.8/3$
\enddata

\tablecomments{(1) Rest-frame energy of Gaussian absorption line; (2) Absorption line equivalent width, for constant width of $\sigma=(240\pm70)$\,eV; (3) \xstar parameters for fit with variable $\lognh$ for a constant $\logxi=3.48\pm0.14$; (4) \xstar parameters for fit with variable $\logxi$, for a constant $\lognh=23.6^{+0.1}_{-0.2}$; (5) outflow velocity in the quasar rest frame ($z=0.184$). Negative values denote a net blue-shift; (6) absorbed (unabsorbed) source flux between $0.5-10$\,keV, in units of $10^{-12}$\flux; (7) change in $\chidof$ when modeling absorption with a Gaussian. Negative values denote an improvement.} 
\tablenotetext{t}{indicates that a parameter is fixed during spectral fitting}
\end{deluxetable*}

\section{Physical properties of the wind}
The observed line-variability can be used to constrain both the properties of the absorbing gas and its location with respect to the continuum source. Here, we use the results from our models to investigate the properties of the high-velocity outflow in PDS\,456.

\subsection{A recombining absorber}
In our first scenario (Model A) we showed that the line variability could be due to an absorber whose ionization parameter decreases from $\logxi>4.28$ to $\sim3.31$, for a constant $\lognh\simeq23.6$. Between when the absorption is first detected (slice 4) and when it reaches its strongest (slice 7) the ionization state of the absorber appears to decrease in proportion to the source flux; $\logxi$ decreases by factor of $\sim3$ (in linear space) from $3.95$ to $3.49$, whilst the absorbed (unabsorbed) $0.5-10$\,keV flux drops by a similar factor during the same time-frame (i.e., from $F_{\rm 0.5-10}=4.72\,(11.81)\times10^{-12}$ in slice 4 to a mean $\sim\langle1.8\,(4.7)\rangle\times10^{-12}$\flux in slices 5--8; see Table~\ref{tab:slices}). This implies that the observed variability could be due to recombination within a smooth (i.e., with constant $\nh$) outflow which is in photo-ionization equilibrium. The gradual decrease in $\xi$ between slices 5--8 would then be due to \fexxvi recombining into \fexxv in delayed response to the bright flare in flux that occurs in slice 4. This scenario could potentially also account for the subtle decrease in velocity shift between slices 5/6 and slices 7/8, with the increasing contribution from \fexxv broadening the profile and giving a lower apparent centroid energy. 

If this is the case, we can estimate the electronic density, $n_{e}$, from the recombination time, $t_{\rm rec}$. The recombination time for ionic population $X_{i}$ depends upon both the rate at which the $X_{i+1}$ ions fall into population $X_{i}$, and the rate at which ions already in the $X_{i}$ population fall into the $X_{i-1}$ population. A robust formula which takes into account these effects is given by \citet{bottorff00}:
\begin{equation}
\footnotesize
 	t_{\rm rec}(X_{i})=\left\{ \alpha(X_{i},T_{e})n_{e}\left[\frac{f(X_{i+1})}{f(X_{i})}-\frac{\alpha(X_{i-1},T_{e})}{\alpha(X_{i},T_{e})}\right]\right\}^{-1},
 	\label{eqn:recomb}
\end{equation}
where $f(X_{i})$ is the fraction of ions in the $X_{i}$ population, $\alpha(X_{i},T_{e})$ is the recombination co-efficient of the $X_{i}$ ion for electron temperature $T_{e}$, and $n_{e}$ is the electron number density. We apply this equation to \fexxv. The appropriate recombination co-efficient is dependent upon the temperature of the gas, which is likely to be of the order $\log(T_{e}/\rm K)\simeq7.7$ for a mean $\logxi=3.7$ (see \citealt{kallman04}, their fig.~6). At this ionization and temperature, the ionic fraction of \fexxvi is roughly twice that of \fexxv, whilst the ratio of Fe\,{\sc xxiv} to \fexxv recombination co-efficient is around unity (see \citealt{nahar01}). Equation (\ref{eqn:recomb}) then reduces to the familiar $t_{\rm rec}(X_{i})\simeq[\alpha(X_{i},T_{e})n_{e}]^{-1}$.

We can estimate the recombination time-scale through the time between the initial onset of the absorption and it reaching maximum depth. Taking the time between the end of slice 4 ($450$\,ks) and the start of slice 7 ($1.25$\,Ms) hence implies $t_{\rm rec}\sim800$\,ks. Using the \fexxv recombination co-efficient from \citet{nahar01} appropriate for the likely electron temperature of $\log(T_{e}/\rm K)\simeq7.7$ (see above), and a recombination time of $t_{\rm rec}\sim800$\,ks, yields $n_{e}=5.5\times10^{5}$\,\pcmcu for the absorbing gas. However, it is important to note that the profile appears to grow gradually, with an ionisation state which decreases with time (see Table~\ref{tab:slices}), meaning that this method will over-estimate the recombination time-scale since it is simply measuring the time it takes for the profile to reach its maximum depth, and not the rate at which it takes the gas to actually recombine. A more physically motivated (and more conservative) way of estimating $t_{\rm rec}$ is to instead consider the ionisation state of the gas directly. A conservative estimate on the recombination time-scale can be obtained through the shortest time it takes for a change in $\xi$ to be discernible in the data. At 90\% confidence, this occurs between slices 6 ($1.15$\,Ms) and slice 8 ($1.35$\,Ms), implying that $t_{\rm rec}\sim200$\,ks, and that hence $n_{e}=2\times10^{6}$\,\pcmcu.

From our estimate on $n_{e}$ we can use the definition of the ionization parameter, $\xi=L_{\rm ion}/n_{\rm H}r^{2}$ (\citealt{tarter69}), to estimate the distance, $r$, of the the absorber from the ionising continuum source. Here, $L_{\rm ion}$ is the integrated ionising luminosity between $1-1000$ Rydbergs. As it is unclear whether or not the partially-covering gas is shielding the outflow we should consider both the absorbed and unabsorbed ionising luminosities, corresponding to the wind being shielded and unshielded by the partial coverer, respectively, to give the likely range in luminosity `seen' by the absorbing material. To first order we can estimate $L_{\rm ion}$ by extrapolating our continuum model into the UV regime. It is worth noting, however, that estimating $L_{\rm ion}$ in this manner will probably lead an under-estimate of the \emph{true} ionising luminosity because it neglects the excess contribution from the UV. 

To roughly gauge how strongly the UV emission is PDS\,456 is likely to contribute to $L_{\rm ion}$ we can compare the flux predicted by our baseline X-ray model to the UV flux reported by \citet{o'brien05}. An extrapolation of the $\Gamma=2.4$ X-ray model predicts a $(1500-2300)\AA$ flux a factor of $\sim3$ lower than observed by the Hubble Space Telescope in the same wavelength range\footnote{although noting the observations are not simultaneous.} (see \citealt{o'brien05}). The discrepancy can be accounted for with a rudimentary break in the UV-to-X-ray Spectral Energy Distribution (SED), hardening from $\Gamma_{\rm UV}\simeq2.7-2.8$ in the UV to $\Gamma_{\rm X}=2.4$ in the X-ray (for an $E_{\rm break}\sim100$\,eV), which increases the total inferred ionising luminosity by a factor of $\sim3$. Even so, because this increase occurs predominantly in the UV regime it will essentially have no effect at Fe\,K where the ionization potential is significantly larger than the typical UV photon energy, i.e., $\sim9$\,keV for \fexxvxxvi. Moreover, while $L_{\rm ion}$ will undoubtedly increase, the value for $\xi$ inferred by \xstar will also increase in proportion, so the overall ratio $L_{\rm ion}/\xi$ remains roughly constant (see fig.~9 in \citealt{giustini11} and discussion therein). Thus, because there is no net effect of accounting for the UV break in these data, we can safely estimate $L_{\rm ion}$ based on our extrapolated X-ray model. This leads to $L_{\rm ion}=(4.0-22.8)\times10^{44}$\,\ergps in the shielded--unshielded cases (corrected for Galactic absorption in both cases).

Thus, for a mean $\logxi=3.7$, and remembering that electron density ($n_{e}$) is related to the gas density ($n_{\rm H}$) by the relation $n_{e}\sim1.2n_{\rm H}$, we have $r=(L_{\rm ion}/n_{\rm H}\xi)^{1/2}\sim(2.2-5.2)\times10^{17}\,\text{cm}$, or $\sim0.07-0.17$\,pc. For a $\mbh\approx10^{9}$\,\msun appropriate for PDS\,456 (R09, R13), this corresponds to a radius of $\sim1500-3500$\,\rg\ from the black hole. This is comparable with inner regions of the Broad Line Region (BLR), at the order of $\sim1000$\,\rg\ (\citealt{o'brien05}, R09) from the black hole. However, we note that this estimate is likely only the characteristic radius for the responding material. Should the outflow be stratified along the LOS (e.g., as in \citealt{tombesi13}) there could still be material over a large extended range of radii.

\subsection{A transiting cloud}
In the second scenario (our Model B) we showed that the absorption could also be modelled with a column density which increases from $\lognh<22.6$ to a maximum $\sim23.8$ over the course of the observation, with a constant $\logxi\sim3.45$. Detailed simulations of accretion disk winds have shown that their ejecta are often clumpy, with a complex density structure (e.g., \citealt{kurosawa09, sim10b, takeuchi13}). If changes in column density are indeed responsible for the observed line variability in PDS\,456, one possibility is that the changes could be associated with an inhomogeneous clump of material within a disk-wind. Alternatively, the clump could originate in an outburst of material being blown from the disk surface during the flaring period. 

Regardless of its origin, the linear extent (diameter) of a transiting clump can be estimated from $\Delta d_{\rm cloud}=v \Delta t$, where $v_{t}$ is the cloud velocity tangential to the LOS and $\Delta t$ is the total duration of the transit. Taking $\Delta t=2\times400$\,ks, corresponding to twice the time between the first definitive detection of the absorption in slice 5 and it reaching its maximum depth in slice 7, and making the reasonable assumption that $v_{\rm t}\sim v_{\rm out}\simeq0.25$\,c, we estimate $\Delta d_{\rm cloud}\approx6\times10^{15}$\,cm (or $\sim40$\,\rg\ in PDS\,456). Taking $\lognh=23.8$ (as measured in slice 7 when the absorption was deepest), and assuming that the absorbing cloud is spherically symmetric, we can then infer its density as $n_{\rm H}=\nh/d_{\rm cloud}\approx1\times10^{8}$\,\pcmcu. This then leads to $r=(L_{\rm ion}/n_{\rm H}\xi)^{1/2}\sim(2.8-6.7)\times10^{16}\,\text{cm}\simeq200-500$\,\rg, depending upon whether the outflow is shielded or not.

Overall, the inferred wind properties are consistent with those posited by R13 after applying the \citet{sim08, sim10a} disk-wind model to the 2007--2011 \suzaku spectra. We also note the similarity between our estimated parameters for the wind and those recently predicted for a time-variable disk-wind by \citet{takeuchi13}. On the basis of R-MHD simulations \citeauthor{takeuchi13} show that a continuum-driven disk-wind can become significantly clumpy or filamentary at $r\gtrsim400$\,\rg\ from the black hole, with the clumps having an optical depth $\tau$ of order unity, an ionization parameter $\xi\sim10^{3}$\,\ergcmps and an observational variability time-scale of roughly one week. These values are similar to those found here for our Model B.

\section{Implications for outflow energetics}
\label{sec:energetics}
Therefore the outflow in PDS\,456 is constrained to lie between $200-3500$\,\rg\ from the black-hole, depending upon the model adopted and whether the gas is shielded. We now consider the energetics associated with such a wind. We first calculate the mass outflow rate using the expression $\dot M_{\rm out}\equiv\Omega m_{p} \nh v_{\rm out} r$, where $\Omega$ is a parameter which sets the overall wind geometry (in terms of the solid angle). This equation is the same as the one derived by \citet{krongold07}, with $\Omega=(6/5)\pi$ being identical to their case of a vertically launched bi-conical wind with solar abundances that has an average angle of $30^{\circ}$ with respect to the LOS (see \citealt{krongold07} for details). Because $\dot M_{\rm out}\propto r$ we consider here only the most conservative estimate of $r\gtrsim200$\,\rg\ as this will lead to a similarly conservative estimate on the wind energetics. As an aside it is also important to note that the two limiting cases that we have considered in this work, i.e., that the wind is either shielded or unshielded from the full ionising continuum by the partially-covering gas, are likely to represent the absolute extreme scenarios. In the former case, should the shielding gas be ionised rather than neutral (as has been assumed here for simplicity) then the reduced gas opacity (and hence enhanced transmission) will allow a larger fraction of $L_{\rm ion}$ to be `seen' by the wind, leading to a larger inferred distance $r$. The lower estimate of $r\simeq200$\,\rg considered here should therefore be regarded as extremely conservative; if a larger estimate on $r$ was adopted instead then the subsequent energetics will be correspondingly larger.

Thus, for a mean $\lognh=23.6$ (taken over slices 4--8) appropriate for our Model B, taking $r\gtrsim200\,r_{\rm g}\simeq2.8\times10^{16}$\,cm, and adopting $\vout=0.25$\,c, we conservatively estimate that $\mout\gtrsim5.3\times10^{26}\,\text{g\,s}^{-1}\simeq8$\msunyr in PDS\,456. This can informatively be written in terms of the Eddington mass accretion rate, $\dot M_{\rm edd}=L_{\rm edd}/\eta c^{2}$, where $L_{\rm edd}=4 \pi G m_{\rm p} \mbh c \sigma_{T}^{-1}\simeq1.26\times10^{38}(\mbh/M_{\odot})$\,\ergps is the Eddington luminosity and $\eta$ is the accretion efficiency of the black hole. For our estimate on the mass outflow rate we find $\mout/\dot M_{\rm edd}\gtrsim0.4(\eta/0.1)(\mbh/10^{9}M_{\odot})^{-1}$, which implies that the mass outflow rate is at least $\sim40\%$ of the Eddington accretion rate for a reasonable $\eta=0.1$. Similarly, we can estimate the wind kinetic luminosity in Eddington units as $L_{\rm kin}/L_{\rm edd}\gtrsim1(\vout/c)^{2}(\mbh/10^{9}M_{\odot})^{-1}$, whilst the momentum rate of the flow is given by $\pout/\pedd\gtrsim4(\vout/c)(\mbh/10^{9}M_{\odot})^{-1}$. For the values appropriate here, i.e., $\mbh\simeq10^{9}$\msun and $\vout\simeq0.25$\,c, we hence estimate $L_{\rm kin}\gtrsim0.06L_{\rm edd}$. This corresponds to $\sim8\times10^{45}$\,\ergps which larger than the typical $\sim0.5-5\%$ of $\lbol$$(\equiv10^{47}$\,\ergps in PDS\,456; \citealt{simpson99, reeves00}) thought necessary for significant feedback (\citealt{dimatteo05,hopkins10}). Moreover, the momentum rate in the flow is comparable to that expected for the Eddington-limited photon field, i.e., $\pout\gtrsim\pedd$, which strongly suggests that the wind was radiatively-accelerated by continuum-scattering processes and/or radiation pressure during an episode of near-Eddington-limited accretion, as has been argued by \citet{king03,king10}.

Ultimately, whilst the overall wind energetics are likely sensitive to the overall flow geometry (e.g., \citealt{giustini12}), these results imply a total integrated energy budget in the wind of $E_{\rm out}\sim10^{60}$\,erg, for a representative quasar active phase of $\sim10^{8}$\,yr with a $\sim10\%$ duty cycle. Thus, even in the {\it most conservative case}, the wind in PDS\,456 could plausibly impart sufficient energy into the host galaxy to exceed the $\sim10^{59}$\,erg binding energy of a mass $10^{11}$\,\msun galaxy bulge (with $\sigma=300$\kms). This supports the argument that the Fe\,K wind in PDS\,456 --- and by extension those also observed in the wider AGN population (e.g., \citealt{tombesi10, tombesi12, gofford13}) --- may play an important role in shaping the host galaxy through feedback. 

\section{Origin of the absorbing gas}
\label{sec:origin_of_the_absorbing_gas}
Through this work we have shown that despite the gradually increasing absorption at $\sim9$\,keV (rest-frame) being a defining characteristic of the last $\sim1$\,Ms of the 2013 \suzaku observation, the same absorption does not appear to be strongly present during the first $\sim400$\,ks. This raises an interesting question: where was the absorption at the start of the observation? We speculate briefly as to the reason for this on the basis of our two different absorption geometries. If the absorption is due to gas in photo-ionization equilibrium one possibility is that it is simply too highly ionised at the start of the observation to be observable in the spectrum. The light-curve (Figure~\ref{fig:lc_slices}) shows that slice~1 traces a factor of $\sim2$ decrease in source flux, perhaps following an even earlier flaring event. If this is the case then the gas, having already been ionised by an earlier flare, may not have had time to sufficiently recombine into an observable feature before subsequently being re-ionised by the strong flare in slice~4. After slice~4, where there are no further strong flares, the gas may then have ample time to recombine, eventually giving rise to the the observed deep feature in slices~7--8.

Alternatively, should the absorption instead be due to a transiting inhomogeneous clump of gas, it is possible that the gas may not be present along the LOS at the start of observation. The onset of the absorption is first apparent in slice~4 which makes it tempting to speculate that the absorbing clump may be in some way related to the flare, perhaps, for example, being blown off of the disk by magnetic flaring in the Comptonizing corona and then further accelerated to high velocity by continuum scattering. Note that the observed outflow velocity of the gas corresponds to an escape radius of $r_{\rm esc}=2(c^{2}/\vout^{2})\simeq32$\,\rg from the black hole. We could therefore be seeing the gas close to where it was launched from the disc, which lends weight to the idea that the absorption could be due to a clump of gas which occults the LOS.

\section{Conclusions}
This paper presents the first results from an extensive \suzaku, \xmm and \nustar observing campaign of powerful quasar PDS\,456, occurring between Feb--Sept 2013. We have reported on the remarkably variable high-velocity Fe\,K-shell wind which is evident in the new long ($\sim1$\,Ms duration) \suzaku observation. Consistent with earlier (2007, 2011) \suzaku observations, the wind is again detected through absorption at $\sim9$\,keV in the source rest-frame ($\vout\sim0.25$\,c). The absorption line depth increase by a factor of $\sim5$ during the new observation. This variability is equally well modelled by (i) an outflow in photo-ionization equilibrium which recombines in response to decreasing source flux, or (ii) an inhomogeneous clump of gas which transits the line-of-sight to the quasar. The variability allows us to directly determine the radius to the gas, e.g., as part of a clumpy outflow, which is constrained to lie between $r\sim10^{2}-10^{3}$\,\rg of the black hole. Even in the most conservative case, the kinetic power of the flow is a significant fraction ($\gtrsim6\%$) of the Eddington luminosity, and is comparable to the $\sim0.5-5\%$ of $\lbol$ thought necessary for significant feedback. The momentum rate of the flow is equivalent to the Eddington momentum rate which is consistent with the flow being radiatively-accelerated by electron scattering during a near-Eddington-limited accretion episode. 

\section*{Acknowledgements}
We thanks the anonymous referee for their comments and suggestions which helped improve the clarity of this manuscript. J.~Gofford, J.~N.~Reeves, G.~A.~Matzeu, and M.~T.~Costa acknowledge financial support from the STFC. J.~N.~Reeves also acknowledges support from \chandra grant number GO1-12143X. This research has made use of data obtained from the \suzaku satellite, a collaborative mission between the space agencies of Japan (JAXA) and the USA (NASA).

\end{document}